\documentclass[fleqn,usenatbib]{mnras} 

\usepackage{newtxtext,newtxmath}

\usepackage[T1]{fontenc}
\usepackage{ae,aecompl}

\usepackage{graphicx}	
\usepackage{amsmath}	
\usepackage{amssymb}	
\usepackage{grffile}    

\def\Mdot{\hbox{${\dot M}$}}

\def\cm{{\rm\thinspace cm}}
\def\km{{\rm\thinspace km}}
\def\s{{\rm\thinspace s}}
\def\yr{{\rm\thinspace yr}}

\def\kmps{\hbox{${\rm\km\s^{-1}\,}$}}

\def\Msol{\hbox{${\rm\thinspace M_{\odot}}$}}

\def\Msolpyr{\hbox{${\rm\Msol\yr^{-1}\,}$}}
\def\spose#1{\hbox to 0pt{#1\hss}}
\def\ltsimm{\mathrel{\spose{\lower 3pt\hbox{$\sim$}}
        \raise 2.0pt\hbox{$<$}}}
\def\gtsimm{\mathrel{\spose{\lower 3pt\hbox{$\sim$}}
        \raise 2.0pt\hbox{$>$}}}

\title[Colliding Stellar Winds]{Colliding Stellar
    Winds Structure and X-ray Emission}

\author[J.~M.~Pittard \& B.~Dawson]
{J.~M.~Pittard\thanks{E-mail: jmp@ast.leeds.ac.uk (JMP)} \& B.~Dawson\\
School of Physics and Astronomy, University of
       Leeds, Woodhouse Lane, Leeds LS2 9JT, UK
}

\date{Accepted 2018 April 13. Received 2018 March 23; in original form
  2017 July 24}

\pubyear{2017}

\begin{document}
\label{firstpage}
\pagerange{\pageref{firstpage}--\pageref{lastpage}}
\maketitle

\begin{abstract}
  We investigate the structure and X-ray emission from the colliding
  stellar winds in massive star binaries. We find that the opening
  angle of the contact discontinuity (CD) is overestimated by several
  formulae in the literature at very small values of the wind momentum
  ratio, $\eta$. We find also that the shocks in the primary
  (dominant) and secondary winds flare by $\approx 20^{\circ}$
  compared to the CD, and that the entire secondary wind is shocked
  when $\eta \ltsimm 0.02$. Analytical expressions for the opening
  angles of the shocks, and the fraction of each wind that is shocked,
  are provided. We find that the X-ray luminosity
  $L_{\rm x} \propto \eta$, and that the spectrum softens slightly as
  $\eta$ decreases.
\end{abstract}




\section{Introduction}
\label{sec:intro}
In binary systems composed of two massive stars, a region of shocked
gas is created if their stellar winds collide. Since the wind speeds
are typically a thousand kilometers per second or more, the shocked
gas may obtain temperatures in excess of $10^{7}\,$K. The resulting
X-ray emission is typically much harder than that from single massive
stars, and may show phase-dependent variability due to changes in the
stellar separation, wind absorption and stellar occultation. Examples
include O+O systems such as Cyg~OB2~No.8A
\citep{DeBecker:2006,Cazorla:2014} and Cyg~OB2~No.9 \citep{Naze:2012},
and WR+O systems such as WR\,11 ($\gamma^{2}$~Velorum)
\citep{Skinner:2001,Schild:2004,Henley:2005}, WR\,21a
\citep{Gosset:2016}, WR\,22 \citep{Gosset:2009}, WR\,25
\citep{Raassen:2003,Pandey:2014}, WR\,139 (V444~Cygni)
\citep{Lomax:2015}, and WR\,140
\citep{Zhekov:2000,Pollock:2005,DeBecker:2011,Sugawara:2015}. Tables
of X-ray luminous O+O and WR binaries were presented by
\citet{Gagne:2012}. The most X-ray luminous binary reported in this
work is WR\,48a, a WC8+WN8h system with an orbital period of about
32\,yrs \citep{Zhekov:2011,Williams:2012,Zhekov:2014}. Perhaps the
best studied system, and certainly one of the most complex, is the
extraordinary LBV-like~+~(WNh?)  binary $\eta$~Car
\citep[e.g.,][]{Corcoran:2001,Corcoran:2005,Hamaguchi:2007,Henley:2008,Corcoran:2010,Hamaguchi:2014a,Hamaguchi:2014b,Hamaguchi:2016,Corcoran:2017}.

Hydrodynamical simulations of the wind-wind collision in massive star
binaries have been presented by many authors
\citep*[e.g.,][]{Luo:1990,Stevens:1992,Myasnikov:1993,Owocki:1995,Pittard:1997,Lemaster:2007,Pittard:2007,Pittard:2009,Lamberts:2011,Parkin:2011a,Parkin:2011b,Parkin:2014,Falceta-Goncalves:2012,Madura:2013,Kissmann:2016}. In
wide systems and/or those with high wind speeds and low mass-loss
rates, the plasma in the wind-wind collision region (WCR) behaves
almost adiabatically, since its cooling time, $t_{\rm cool}$, is much
greater than the time it takes to flow out of the system,
$t_{\rm esc}$. \citet{Stevens:1992} introduced a cooling parameter,
$\chi$, which is the ratio of these timescales
($\chi = t_{\rm cool}/t_{\rm esc}$). In systems with $\chi >> 1$ the
gas in the WCR behaves almost adiabatically, while in those with
$\chi \ltsimm 1$ radiative cooling effects are
important. \citet{Stevens:1992} showed that the nature of the WCR, and
the instabilities that it may experience, are closely tied to the
value of $\chi$ for each of the winds. In adiabatic systems strong
instabilities are largely absent, though the Kelvin Helmholtz
instability may be present if the velocity shear at the contact
discontinuity which separates the winds is significant. In contrast,
systems where both the shocked primary and secondary winds strongly
cool are susceptible to thin-shell instabilities which disrupt and
``shred'' the thin shell, creating violent, large-amplitude
oscillations in the process
\citep{Stevens:1992,Kee:2014,Pittard:2018}.

\citet{Stevens:1992} showed that the X-ray emission from adiabatic
systems scales as the inverse of the stellar separation (i.e.
$L_{\rm x} \propto 1/D_{\rm sep}$). Assuming that the emitting volume
of the WCR scales as the cube of the distance from the weaker star to
the stagnation point, $d_{2}^{3}$, they further noted that the
adiabatic luminosity should scale as
$(1 + \mathcal{R})/\mathcal{R}^{4}$ (their Eq.~10), where
$\mathcal{R} = (\Mdot_{1}v_{1}/\Mdot_{2}v_{2})^{1/2}$, the mass-loss
rates are $\Mdot_{1}$ and $\Mdot_{2}$, and the wind speeds are $v_{1}$
and $v_{2}$\footnote{Subscript 1 indicates quantities measured for the primary
star, and subscript 2 indicates those measured for the secondary.  In
all of the following we will refer to the star with the stronger wind
as the ``primary'' star, and to the star with the weaker wind as the
``secondary'' star.}. The wind momentum ratio is usually defined in the
literature as $\eta$, such that
$\eta = \Mdot_{2}v_{2}/\Mdot_{1}v_{1}$. Thus
$\mathcal{R} = \sqrt{1/\eta}$.

\citet{Pittard:2002} showed that for systems with equal wind speeds
and identical compositions, the dominant wind is also the dominant
X-ray emitter (see their Table~3). For instance, when $\eta=0.01$, the
X-ray emission from the shocked primary wind is $24\times$ greater
than that from the shocked secondary wind (despite a greater
proportion of the secondary wind being shocked). This is due to the
fact that the stronger wind becomes more efficient at radiating
relative to the weaker wind (the ratio of the cooling parameter for
the two winds is
$\chi_{1}/\chi_{2} \sim \dot{M}_{2}v_{1}^{4}/\dot{M}_{1}v_{2}^{4}$ -
see \citet{Pittard:2002} for further details).

To our knowledge, the scaling of $L_{\rm x}$ with
$\mathcal{R}$ proposed by \citet{Stevens:1992} has never been tested, yet it
is fundamental to some analyses in the literature
\citep[e.g.,][]{Sugawara:2015}. Therefore we investigate this scaling
in this paper, along with the opening angles of the CD and each wind's
shock.

\section{The Numerics}
\label{sec:numerics}
The structure of the WCR is calculated using a hydrodynamics code
which is 2nd order accurate in space and time. The code solves the
Euler equations of inviscid fluid flow on a 2D axisymmetric grid. The
cell-averaged fluid variables are linearly interpolated to obtain the
face-centered values which are input to a Riemann solver. A linear solver
is used in most instances, but a non-linear solver is used when the
difference between the two states is large \citep{Falle:1991}. The
solution is first evolved by half a time step, at which point
fluxes are calculated with which to advance the initial solution by a
full time step. A small amount of artificial viscosity is added to the
code to damp numerical instabilities. All calculations were performed
for an adiabatic, ideal gas with $\gamma=5/3$. The pre-shock wind
temperature is kept constant at $10^{4}\,$K.

The grid has a reflecting boundary on the $r=0$ axis. All other
boundaries are set to enable outflow. The stellar winds are mapped
onto the grid at the start of every time step by resetting the
density, pressure and velocity values within a region of 10-cell
radius around each wind. To avoid any axis effects, care is taken to
use the position of the cell centre-of-mass when calculating these
values, and also when linearly interpolating the fluid variables for
input to the Riemann solver and when calculating the source
term in the r-momentum equation \citep[see][for further
details]{Falle:1991}.

The initial conditions are of two spherically expanding winds
separated by a planar discontinuity which passes through the
stagnation point of the wind-wind collision. The solution is then
evolved for many flow timescales until all initial conditions have
propagated off the grid and the solution has reached a stationary
state. Typical calculations use a grid of $\sim 10^{6}$ cells, though
extremely low values of $\eta$ require substantially more.

Our standard simulation has $\dot{M}_{1,2} = 10^{-6}\,\Msolpyr$,
$v_{1,2} = 2000\,\kmps$, and $D_{\rm sep}=10^{14}\,\cm$. The wind
parameters are typical for massive stars while the adopted separation
means that the wind acceleration can be ignored (i.e. the winds are
assumed to collide at their terminal speeds).  The mass-loss rate of
the secondary star, $\dot{M}_{2}$, is reduced to study the wind-wind
interaction in systems with unequal strength winds.

\begin{figure*}
\includegraphics[width=5.5cm]{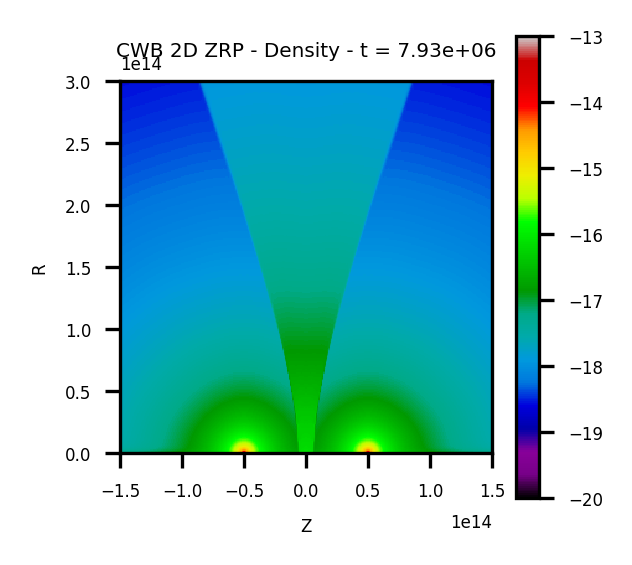}
\includegraphics[width=5.5cm]{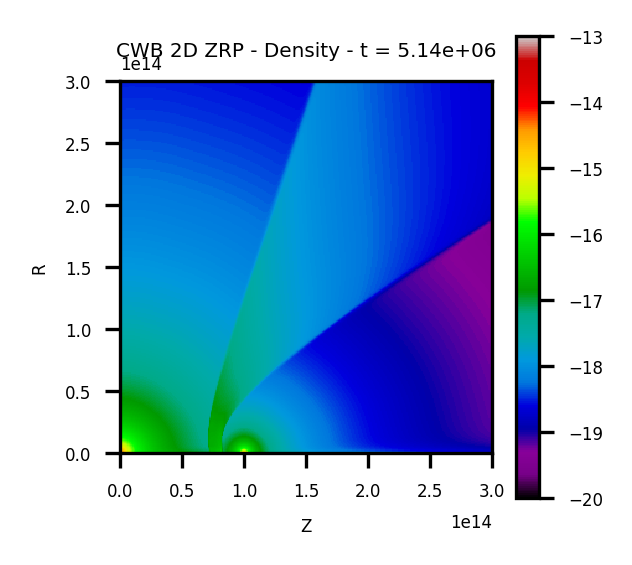}
\includegraphics[width=5.5cm]{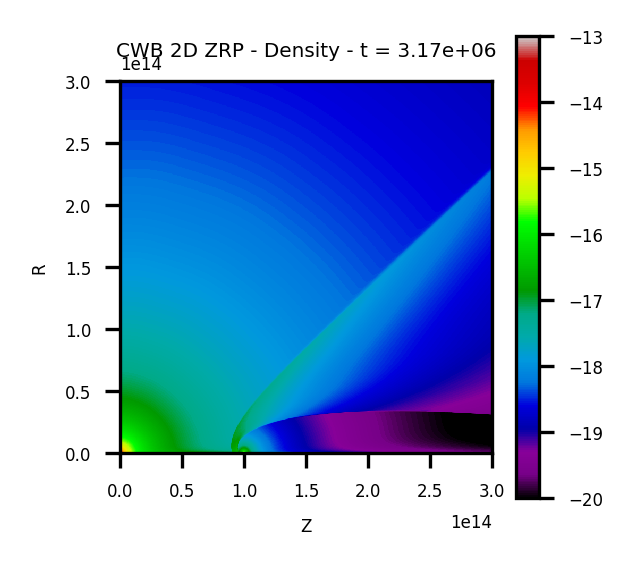}
\caption{Density distributions in the winds and WCR for 3
  different values of $\eta$: $\eta=1.0$ (left), $\eta=0.1$ (middle),
  $\eta=0.01$ (right). The primary wind parameters and $v_{2}$ were kept fixed, while
  $\dot{M}_{2}$ was varied. Larger grids than shown were used to calculate the
  X-ray emission when $\eta < 1$. Distances are in cm, and densities
  in ${\rm g\,cm^{-3}}$.}
\label{fig:dens}
\end{figure*}

Systems with an equal wind momentum ratio, $\eta=1$, produce a wind
collision region which is symmetric and equi-distant from the
stars. The contact discontinuity is a plane and the reverse shocks
bend towards each star. In this case $(1/6)^{\rm th}$ of each wind's
kinetic power is thermalized and the X-ray luminosity is
maximal. In systems where one wind is stronger (i.e. has a greater
momentum flux) than the other ($\eta < 1$), the collision region occurs
closer to the secondary star, and forms a
``cone'' around it. In such cases a greater percentage of the
secondary wind passes through the collision region, and a lower
fraction of the primary's. For extreme wind momentum ratios
(i.e. $\eta \ltsimm 0.01$), the collision region becomes so bent over
that \emph{all} of the secondary's wind may be shocked. 
Fig.~\ref{fig:dens} shows the density distribution from 3 models with
different values of $\eta$.

The results of the hydrodynamic calculations are fed into an X-ray emission
code. The X-ray emissivity is calculated using the mekal emission code
\citep{Mewe:1995}, for an optically thin thermal plasma in collisional
ionization equilibrium. Solar abundances \citep{Anders:1989} are
assumed throughout this paper. The emissivity is stored in look-up
tables containing 200 logarithmic energy bins between 0.1 and 10 keV,
and 91 logarithmic temperature bins between $10^{4}$ and
$10^{9}\,$K. Line emission dominates the cooling at temperatures below
$10^{7}\,$K, with thermal bremsstrahlung dominating at higher
temperatures. The hydrodynamical grid is set large enough to capture the
majority of the X-ray emission from each of the models. Since we are
only interested in the intrinsic X-ray emission we do not concern
ourselves with details of the X-ray absorption.

\section{Results}
\label{sec:results} 

\subsection{Opening angles and wind fractions}
\label{sec:angles}
\begin{figure*}
\includegraphics[width=8.3cm]{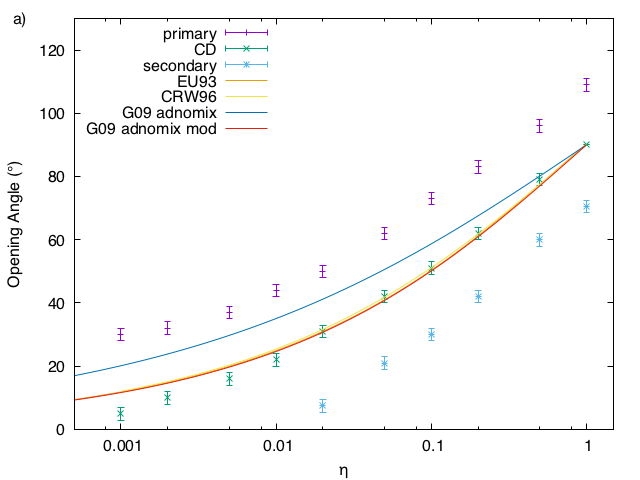}
\includegraphics[width=8.3cm]{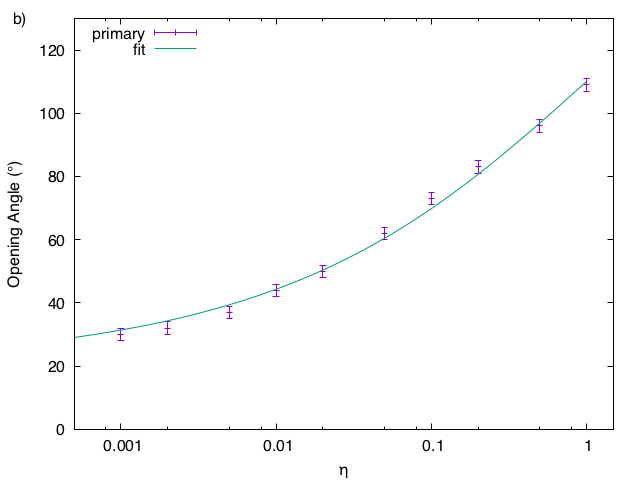}
\includegraphics[width=8.3cm]{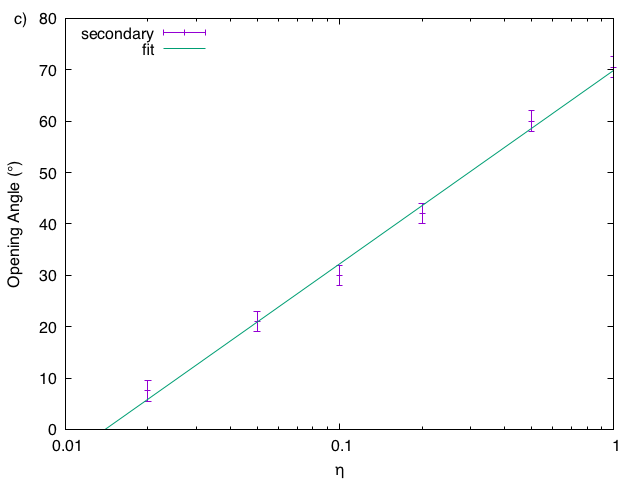}
\includegraphics[width=8.3cm]{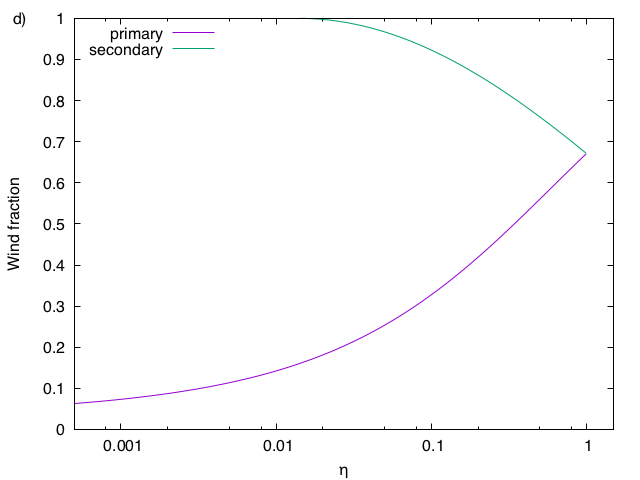}
\caption{a) Opening angles of the primary and secondary shocks, and
  the CD. b) Opening angle of the primary shock. c) Opening angle of
  the secondary shock. d) Fraction of wind shocked.}
\label{fig:openingAngles}
\end{figure*}

Prior to studying the X-ray emission from our simulations we first
examine how the opening angles of the shocks and CD vary with
$\eta$. Here the opening angle, $\theta$, is defined as the angle
between the secondary star, the stagnation point, and the shock or CD.
Table~\ref{tab:openingAngles} and Fig.~\ref{fig:openingAngles}
highlight our findings.

\begin{table}
\centering
\caption[]{The opening angles of the primary ($\theta_{1}$) and secondary ($\theta_{2}$) shocks, and
  the contact discontinuity ($\theta_{\rm CD}$), as a function of $\eta$. All values have an
  estimated uncertainty of $\pm 2^\circ$, except the opening angle of the CD when
  $\eta=1.0$ which is by defintion known to be precisely
  $90^{\circ}$. The entirety of the secondary wind is shocked when $\eta
  \ltsimm 0.01$, so the secondary shock does not have an asymptotic
  opening angle in such cases.}
\label{tab:openingAngles}
\begin{tabular}{lrrr}
\hline
$\eta$ & $\theta_{1}$ & $\theta_{\rm CD}$ & $\theta_{2}$ \\
\hline
1.0 &   109 &	90 &	71 \\	
0.5 &	96 &	79 &	60 \\
0.2 &	83 &	62 &	42 \\ 
0.1 &	73 &	51 &	30 \\
0.05 &	62 &	42 &	21 \\
0.02 &	50 &	31 &	7 \\
0.01 &	44 &	22 &	\\	
0.005 &37 &	16 &	\\
0.002 & 32 & 10 & \\
0.001 & 30 & 5 & \\
\hline
\end{tabular}
\end{table}

These values are in good agreement with an earlier determination from
hydrodynamical simulations \citep{Pittard:2006}. When $\eta=1$ the winds are of equal strength,
and the shocks flare out by $\approx 19^{\circ}$ from the CD. The
secondary shock has $\theta_{2} = 0.0$ (i.e. the secondary wind is
completely shocked) when $\eta$ is just above 0.01 (at $\eta=0.01$ the
secondary shock is curving back towards the line of symmetry).

As far as we are aware, there are no other measurements of the shock
and CD opening angles from hydrodynamical simulations in the
literature\footnote{\citet{Lamberts:2011} report on the
    positions of the shocks and the CD in 2D calculations.}. However,
there have been numerous attempts to determine analytical expressions
for the opening angles of the CD. For instance, \citet{Girard:1987}
assumed that the shocks were highly radiative (and thus spatially
coincident with the CD), and calculated their position based on
momentum conservation.
\citet{Eichler:1993} report that \citet{Girard:1987}'s results are
well approximated by the function
\begin{equation}
\label{eq:eu93}
\theta \approx 2.1 \left(1 - \frac{\eta^{2/5}}{4}\right) \eta^{1/3}.
\end{equation}
\citet{Canto:1996} also investigated the case of highly radiative
shocks, and provided a formula for the opening angle (albeit using a
different definition for $\theta$). Changing to the
usual definition one finds that their Eq.~28 is equivalent to
\begin{equation}
\label{eq:crw96}
\theta - \tan\theta = \frac{\pi\eta}{\eta-1}.
\end{equation}
More recently, the ``characteristic'' opening angle of an adiabatic wind-wind collision
was considered by \citet{Gayley:2009}. As a result of the shock
heating, an increase momentum flux is generated away from the axis,
and leads to a greater opening
angle than for the case of a radiative WCR. If there is no mixing
across the CD, \citet{Gayley:2009} finds that
\begin{equation}
\label{eq:g09}
\theta = 2\tan^{-1}(\eta^{1/4}).
\end{equation}

Fig.~\ref{fig:openingAngles}a) shows the functions in
Eqs.~\ref{eq:eu93}-\ref{eq:g09} plotted against our results. We find
that the \citet{Eichler:1993} and \citep{Canto:1996} formulae are
almost identical, while the \citet{Gayley:2009} formula produces
larger opening angles for $\eta < 1$. We also find that modifying
Gayley's formula to $\theta = 2\tan^{-1}(\eta^{1/3})$ brings it back
into agreement with the other formulae. This is also consistent with
the discusion in Sec.~4.1 in \citet{Gayley:2009}.  We further note
that while the \citet{Eichler:1993}, \citet{Canto:1996} and our
``modified'' Gayley formulae fit the results from our hydrodynamical
simulations very well for $0.01 \ltsimm \eta \ltsimm 1$, the opening
angle becomes increasingly divergent at smaller values of $\eta$.

In contrast to the many functions which exist for $\theta_{\rm CD}$,
there are no formulae for the opening angles of the shocks,
$\theta_{1}$ and $\theta_{2}$, when the WCR is not highly
radiative\footnote{\citet{Usov:1992} provides an
  expression for the position of the primary shock when the primary
  wind completely overwhelms the secondary wind and collides directly
  with the secondary star.}. As a matter of interest we note that
multiplying Eq.~\ref{eq:g09} by a factor of $\approx 1.2$ yields a
reasonable fit to $\theta_{1}$, but the opening angle is
underestimated when $\eta \ltsimm 0.005$.  Nevertheless, this shows
that \citet{Gayley:2009}'s ``characteristic'' opening angle perhaps
better describes $\theta_{1}$ than $\theta_{\rm CD}$. We also notice
that the primary shock maintains a roughly constant angle from the CD
as a function of $\eta$. Fig.~\ref{fig:openingAngles}b) shows a fit to
the primary shock position, assuming that
$\theta_{1} = 2\tan^{-1}(\eta^{1/3}) + \delta\theta$. The best fit has
$\delta \theta \approx \pi/9$. Our hydrodynamical simulations do not
extend to $\eta < 10^{-3}$, so we cannot test whether the primary
shock will always achieve an opening angle of at least $20^{\circ}$,
as is implied by this function.
 
Considering now the opening angle of the secondary shock we find that
it is reasonably well fit by the function
$\theta_{2} = 0.658 \log_{10}(71.7\eta)$, which implies that the
entire secondary wind is shocked when $\eta < 1/71.7 \approx 0.014$
(Fig.~\ref{fig:openingAngles}c).

Using our approximations for $\theta_{1}$ and $\theta_{2}$, we can
determine the fraction of each wind which is shocked as a function of
$\eta$. For the primary shock this is $f_{1} = \Omega_{1}/(4\pi)$,
where $\Omega_{1} = 2\pi(1 - \cos\theta_{1})$. For the secondary wind,
this fraction is $f_{2} = \Omega_{2}/(4\pi)$, where
$\Omega_{2} = 2\pi(1 + \cos\theta_{2})$. The resulting fractions are
shown in Fig.~\ref{fig:openingAngles}d).

There appear to be two cases in the literature where the opening
angles are incorrectly calculated.  \citet{Zabalza:2011} estimated
that for small values of $\eta$, $\theta_{\rm CD} \approx \pi\eta$ and
$f_{\rm CD} \approx \theta_{\rm CD}^{2}/4 \approx (\pi \eta)^{2}/4$.
However, for small values of $\eta$, the \citet{Canto:1996} analysis
actually gives
$\theta_{\rm CD}^{3} \approx 3\pi\eta/(1-\eta) \approx 3\pi\eta$
\citep[cf. Sec.~4.1 in][]{Gayley:2009}, which yields
$f_{\rm CD} \approx (3\pi\eta)^{2/3}/4$, rather than the expression
given by \citet{Zabalza:2011}. In any case,
Fig.~\ref{fig:openingAngles}a) shows that the \citet{Canto:1996}
analysis overestimates $\theta_{\rm CD}$ at low values of $\eta$.  The
second occurence is in \citet{Lomax:2015}, where it is noted in
Sec.~4.3 that \citet{Canto:1996}'s formula gives an opening angle of
$68^{\circ}$ for $\eta=0.058$. In fact, it gives
$\theta = 43.7^{\circ}$, in agreement with the other formulations in
their section.

\subsection{The X-ray luminosity and spectral shape}
\label{sec:lx}

\begin{figure*}
\includegraphics[width=8.3cm]{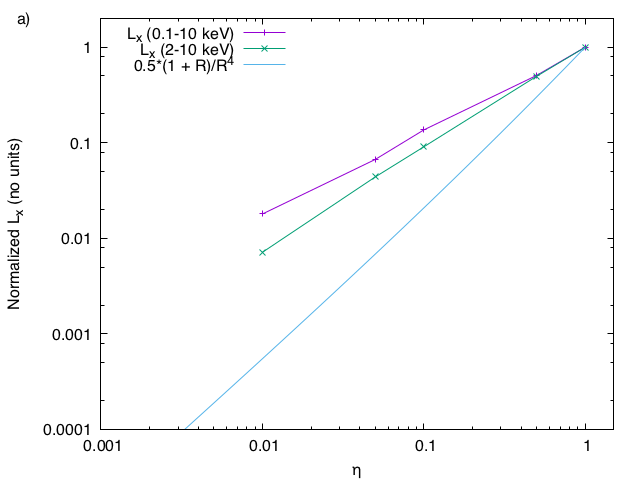}
\includegraphics[width=8.3cm]{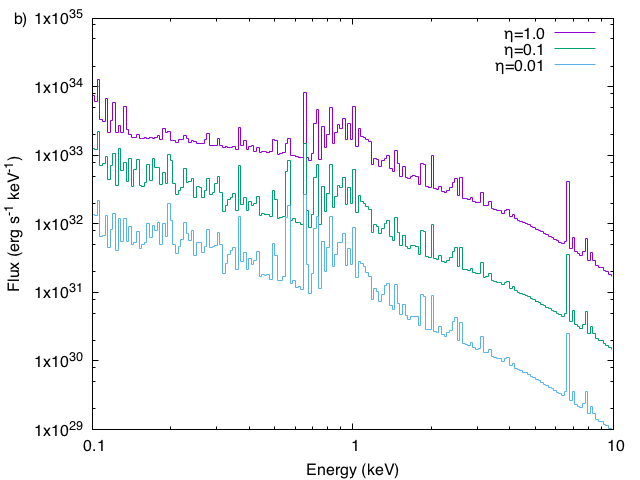}
\caption{a) Scaling of the X-ray luminosity with $\eta$. Also compared
  is the scaling suggested by \citet{Stevens:1992}. b) Variation of
  the X-ray spectrum with $\eta$. All calculations were based on our
  standard model parameters, with $\dot{M}_{2}$ varying.}
\label{fig:lxEta}
\end{figure*}

Fig.~\ref{fig:lxEta}a) shows how the X-ray luminosity calculated from
our hydrodynamical simulations scales with $\eta$. It is immediately
clear that the proposed scaling by \citet{Stevens:1992} is not a good
match to the actual variation in $L_{\rm x}$. The scaling suggested by
\citet{Stevens:1992} goes approximately as
$L_{\rm x} \propto \eta^{3/2}$ (at small values of $\eta$), whereas
the numerical simulations instead scale approximately as
$L_{\rm x} \propto \eta$. Since the distance between the stagnation
point and the secondary star,
$r_{\rm OB} = \sqrt{\eta}D_{\rm sep}/(1 + \sqrt{\eta})$, scales as
$\sqrt{\eta}$ for small $\eta$, this implies that
$L_{\rm x} \propto r_{\rm OB}^{2}$ when $\eta$ is small, which is
akin to the X-ray luminosity scaling in proportion to a ``target
area'' rather than a ``characteristic'' volume.

Fig.~\ref{fig:lxEta}a) shows that the exact variation of $L_{\rm x}$
with $\eta$ also depends on the X-ray band concerned, with the
variation being slightly stronger in harder bands. This is a result of
the shape of the spectrum also being dependent on $\eta$, as shown in
Fig.~\ref{fig:lxEta}b).  Previous work claimed that the spectral shape
was insensitive to the value of $\eta$ \citep{Pittard:2002a}, at least
over the range $1.26-10$\,keV. However, by examining the spectrum over
a greater energy range we now see that there is indeed a small effect. In
particular, we see a change in the slope of the continuum, and changes
to the strength of the line emission, especially for lines below
1\,keV.  That the spectrum softens with decreasing $\eta$ is likely
caused by the increasing dominance of the shocked primary wind to the
X-ray emission, and the increasing obliquity of the primary wind
shock, with decreasing $\eta$. We find this to be the case for
simulations with other wind speeds too (e.g., both winds blowing at
$1500\,\kmps$ or $3000\,\kmps$).

\section{Conclusions}
\label{sec:conclusion}
We have investigated the structure of and X-ray emission from the
colliding stellar winds in massive star binaries. We find that the
opening angle of the CD is in good agreement with previous studies for
$\eta \gtsimm 0.01$, but that these studies overestimate it when
$\eta \ltsimm 0.01$. We also find that the shocks in the primary and
secondary winds flare by about $20^{\circ}$ relative to the CD, and
that this is approximately independent of $\eta$. This implies that
the opening angle of the primary shock does not tend to zero in the
limit $\eta \rightarrow 0.0$. We also find that the X-ray luminosity
scales roughly as $L_{\rm x} \propto \eta$, which is not as steep a
dependence on $\eta$ as previously conjectured, and that the X-ray
spectrum softens slightly as $\eta$ decreases.

It would be very interesting to compare our new predicted scaling
($L_{\rm x} \propto \eta$) with observations. A direct comparison
would require observations of systems where one (or both) winds has
changed in strength. Such systems do exist. For example, the most
massive and luminous binary system in the Small Magellanic Cloud,
HD\,5980, contains a star which underwent an eruptive event in 1994
\citep{Barba:1995}, during which its mass-loss rate increased while
its terminal wind speed decreased. The star has now evolved back
towards something like its pre-eruption state
\citep[e.g.,][]{Foellmi:2008,Georgiev:2011}.  Earlier X-ray
observations revealed orbital phase-dependent variability, but very
recently longer-term changes to the X-ray emission, believed to be due
to the changes in wind properties of the eruptive component, have been
reported \citep{Naze:2018}. Thus HD\,5980 would seem to be the perfect
system against which to test our new predictions. Unfortunately, the
WCR in HD\,5980 is expected to be strongly radiative, even when the
stars are at apastron, whereas our theoretical predictions are for
systems where the WCR behaves largely adiabatically. In future one may
hope to find a system similar to HD\,5980, but where the WCR behaves
adiabatically.

An alternative approach would be to make an indirect comparison to
observations, whereby the observed X-ray luminosity from many systems
is examined. The most straightforward comparison would involve finding
systems where only one of the key parameters changes between them
(e.g., the mass-loss rate of the secondary star), while all others are
comparable (e.g., the wind speeds and stellar separations remain
similar). This task is likely to be difficult, since it will require
accurate measurements of these parameters. Relaxing these requirements
would yield more potential systems, which would perhaps allow a more
indirect, statistical study.

Our simulations were axisymmetric, and ignored details such as the
radiative driving of the winds, orbital motion, radiative cooling of
the shocked gas, and effects such as non-equilibrium ionization,
non-equilibration of electron and ion temperatures, and particle
acceleration, all of which will affect either the shock positions and
structure of the WCR or the resulting X-ray emission. Some or all of
these complications may need to be considered when specific systems
are modelled. However, we hope that our results will be a useful guide
to the analysis and interpretation of systems with colliding winds.

\section*{Acknowledgements}
The calculations for this paper were performed on the DiRAC Facility
jointly funded by STFC, the Large Facilities Capital Fund of BIS and
the University of Leeds. We thank the Royal Astronomical Society for
funding a 6 week summer placement, and the referee for a timely and
useful report. Data for the figures in this paper are available from
\url{https://doi.org/10.5518/349}.






\bsp	
\label{lastpage}
\end{document}